\documentclass[%
superscriptaddress,
preprint,
 amsmath,amssymb,
]{revtex4-2}

\usepackage{graphicx}
\graphicspath{{./Figures/}}
\usepackage{dcolumn}
\usepackage{bm}
\usepackage{psfrag}
\usepackage[percent]{overpic}
\usepackage{amsmath}
\usepackage[caption=false,labelfont={it}]{subfig}
\usepackage[normalem]{ulem}
\usepackage[dvipsnames]{xcolor}
\usepackage{floatrow}
\usepackage{mathtools}
\usepackage{rotating}
\usepackage{mathrsfs}
\usepackage{upgreek}
\usepackage{bm}
\usepackage{tikz}
\usetikzlibrary{calc,patterns,angles,quotes,arrows}
\usepackage{comment}
\usepackage{siunitx}
\def\Oh{{\text{Oh}}}
\def\Ma{{M_{\alpha}}}
\def\Mb{{M_{\beta}}}
\def\E{{\rm \epsilon}}
\def\ubar{\overline{u}}
\def\Ubar{{\overline{U}}}
\def\L{{\mathcal{L}}}
\def\H{{\mathcal{H}}}

\def\dpopa#1#2{{\partial_{#1}#2}}

\definecolor{amber}{rgb}{1.0, 0.49, 0.0}
\def\amb {\color{black}}
\begin{document}

\preprint{APS/123-QED}

\title{Pinching Dynamics of Thin Films of Binary Mixtures}


\author{A. Choudhury}
\affiliation{
PMMH, CNRS, ESPCI Paris, Université PSL, Sorbonne Université, Université de Paris, F-75005, Paris, France
}
\affiliation{
CNRS, Sciences et Ingénierie de la Matière Molle, ESPCI Paris, Université PSL, Sorbonne Université, 75005 Paris, France
}%
\author{L. Duchemin}%
\affiliation{
PMMH, CNRS, ESPCI Paris, Université PSL, Sorbonne Université, Université de Paris, F-75005, Paris, France
}

\author{F. Lequeux}
\affiliation{
CNRS, Sciences et Ingénierie de la Matière Molle, ESPCI Paris, Université PSL, Sorbonne Université, 75005 Paris, France
}
\author{L. Talini}
\affiliation{%
 CNRS, Surface du Verre et Interfaces, Saint-Gobain, 93300 Aubervilliers, France 
}%


\date{\today}

\begin{abstract}
In binary mixtures, the lifetimes of surface bubbles can be five orders of magnitude longer than those in pure liquids because of slightly different compositions of the bulk and the surfaces, leading to a thickness-dependent surface tension of thin films. Taking profit of the resulting simple surface rheology, we derive the equations describing the thickness, flow velocity and surface tension of a single liquid film. Numerical resolution shows that, after a first step of tension equilibration, a parabolic flow with mobile interfaces is associated with film pinching in a further drainage step. Our model paves the way for a better understanding of the rupture dynamics of liquid films.
\end{abstract}

\maketitle

The lifetimes of foams and surface bubbles are primarily governed by the drainage and rupture processes of the thin liquid films surrounding bubbles. These processes are influenced by various intricate - and sometimes difficult to control - factors, including contamination \cite{lhuissier2012bursting}, physicochemical properties of added surfactants \citep{petkova_foamability_2020}, evaporation \cite{pasquet_impact_2022} and history of thin film formation \citep{klaseboer2000film, zawala_coalescence_2023}, which may induce variations in lifetimes over several orders of magnitude. In the presence of surface active species, a comprehensive description of the drainage of thin films is made difficult even in controlled conditions, particularly because of the coupling of flow with concentration field of species. Further complexity arises from the timescales of surface and bulk exchanges of surfactant, which can be comparable to the drainage time \cite{petkova_foamability_2021}. Generally, surface rheology is accounted for by using a velocity boundary condition corresponding to an immobile interface with air, leading to a Poiseuille flow within the film. Although this is a fair assumption leading to a simple lubrication equation for film thickness, it ignores the evolution of surface tension field in the liquid film. 

Recently, the foaming of oil mixtures has attracted renewed interest \citep{chandran_suja_evaporation-induced_2018, lombardi_nonaqueous_2023}. The relatively stable foams that can form in these mixtures without any surfactant have been evidenced decades ago \citep{ross_foaminess_1975}, and are currently observed in many processes of the oil industry, for example car tank filling and crude oil extraction, or in the food industry, for instance, in frying oils. Anti-foaming molecules are then often required to increase efficiency \citep{pugh1996foaming}. In the absence of evaporation, it has been shown that the enhanced lifetimes of thin films in mixtures stem from slight differences between the bulk and surface concentrations of different species, leading to thickness-dependent surface tensions of thin films \citep{tran2020understanding, tran2022mechanisms}. Since the diffusion times of small molecules are very short, bulk and interfaces can be considered to be in thermodyamical equilibrium (the time for a molecule to travel across the thickness of a film is $h^2/D\sim 1~m$s for a $1~\mu$m thick film). Thus, the surface rheology of mixtures reduces to Gibb’s elasticity \citep{tempelapplication1965}. In addition, the disjoining pressure is purely attractive, therefore binary oil mixtures constitute much simpler systems than surfactant solutions to study the drainage mechanisms of liquid films.

Recent insight has been offered on the process of film drainage and rupture, which may be divided into three stages \citep{lhuissier2012bursting,tregouet_instability_2021, tran2022mechanisms}. The first stage comprises of the thin film formation; from spherical surfaces to locally flat surfaces. In case of binary mixtures, this shape can be described using a mechanism of equilibration of film tension by a balance between surface tension gradient and the pressure gradient due to Laplace pressure difference \citep{tran2022mechanisms}. 
This equilibrium shape is reached in a few milliseconds. Naturally, a second stage dynamics ensues, relaxing the pressure gradient and causing a more complicated film drainage scenario. The relaxation causes the film to drain via dimpled (marginal) pinching, as described in soap films \citep{aradian2001marginal, tregouet_instability_2021}. At one point, the film becomes so thin that a third stage of van der Waals attraction becomes effective and causes spontaneous rupture \citep{shah2021influence, liu2023nanoscale}. The film lifetime is thus mostly determined by the second stage of film drainage when a pinched part reaches a critical thickness, which is much longer than the initial viscous stretching phase and the final spontaneous rupture due to van der Waals interactions. Existing models developed for pinching of films of surfactant solutions are based on immobile interfaces \citep{klaseboer2000film,aradian2001marginal,shah2021influence}, and they cannot predict quantitative lifetimes observed in thin films of binary mixtures \cite{tran2022mechanisms}.

In this letter, we formulate a robust model describing the first two stages of marginal pinching of films of binary mixtures. Our model presented here does not impose immobile interfaces a priori, and the pinching dynamics is described by three 1D coupled evolution equations for film thickness $h$, mean flow velocity $\ubar$ and surface tension $\gamma$. In the past, models for films with surfactant-like effects describing parabolic flow with mobile interfaces have been developed \cite{breward2002drainage,liu2023nanoscale}. 
However, they have a general description for surface tension evolution in terms of surfactant concentration with empirical Marangoni parameters. In contrast, we obtain an evolution equation for surface tension, here in the context of binary mixtures, using thermodynamic principles of ideal solution theory. As a result, the parameters defining the surface tension gradient are fully determined by the chosen bulk concentration of the mixture species with lower surface tension. In general, if physicochemical relations between surface tension and the source field of surface elasticity are known (surface excess is well defined), our derivation approach can be readily used to derive the evolution equation of surface tension in other systems like sparse surfactants \citep{lhuissier2012bursting}, electrolyte solutions \citep{liu2023nanoscale}, lipid molecules on biological films \citep{choudhury2021tear,dey2020model} and films of nematic liquid crystals \citep{loudet2022particle} - all governed by linear variation of surface tension.

To test our model, we take a simple geometry of ligament bounded within a characteristic length $\L$, and thickness at the boundaries $\H$. This configuration is close to a film in a Scheludko cell. 
Initially, a flat-film profile of thickness, $h_i=500~n$m is prescribed in the middle (spanning a length of $\mathcal{L}/2$) conjoined to a Plateau border of radius, $R_b=1$ mm. This is described in details in \cite{supplementary}. The estimations are taken from experimental values reported in \cite{tran2022mechanisms}. We model the ensuing drainage dynamics using conservative laws for mass, momentum  and mixture species. In the following, we use subscripts $(x,t)$ to denote derivatives $(\partial_x, \partial_t)$ in the equations. The horizontal velocity is written as $u=\Bar{u}+u_{Po}(z^2/h^2-1/12)+ O(z^4)$. The average velocity $\Bar{u}$ is responsible for the shape evolution of the film, and the velocity $u_{Po}$ - which is proportional to the square root of the variance of the velocity over the vertical direction - is responsible only for the relative motion of the surface and the bulk and thus for the {\amb non-uniform} advection of species. The mass conservation of the film writes as:
\begin{equation}
       D_t h = -h\ubar_x,\label{eq:1}
\end{equation}
where the material derivative $D_t=\partial_t + \ubar\partial_x\sim\partial_t$ since inertia is negligible. The conservation of momentum in a thin liquid film along with interfacial stress conditions can be effectively reduced to an evolution equation for $\ubar$ (see Refs. \citep{supplementary} \& \cite{eggers1994drop} for details):
\begin{equation}\label{eq:2}
D_t\ubar = \frac{1}{\rho h}\left(2\gamma_x+\frac{\gamma_0hh_{xxx}}{2}\right) + \frac{\mu}{\rho}\frac{(4h \ubar_x)_x}{h}.
\end{equation}
Here, $\rho$, $\mu$ are the density and viscosity of the liquid mixture respectively and $\gamma_0$ is the reference surface tension of the liquid mixture at infinite film thickness. The first term in eq.~\eqref{eq:2} can be identified as the gradient of film tension $C$ at the leading order in $h_x$. More precisely, in the limit of small curvatures and slopes, it can be written as $C~=~\gamma \left(2- h_x^2/4 + h h_{xx}/2  \right)$, as obtained previously in Ref. \cite{tran2022mechanisms}. The last term originates from the classical Trouton viscosity which is the ratio of elongational to shear viscosity for planar Newtonian viscous flows appropriate for thin films \citep{choudhury2020enhanced}.

Lastly, we need to introduce the relation between species concentration and surface tension, to couple the flow with the surface tension gradient. 
As discussed in \cite{tran2020understanding, tran2022mechanisms}, under stretching, a film ligament of binary mixture experiences an increase of surface tension. This increase also corresponds to the Gibbs elastic modulus and is given by $\alpha \gamma_0 /h$, where $\alpha$ is a length related to the concentration differences between bulk and interfaces. These differences are modeled in the framework of the thermodynamics of ideal solutions \citep{butler_thermodynamics_1932}, and we assume similar molar volume and surface of both constituents for the sake of simplicity. 
The resulting length $\alpha$ exhibits a maximum value of the order of $10^{-1}~n$m, is a function of the species volume fraction, and vanishes for pure liquids. Since the surface tension varies weakly during the whole pinching process (typically differences of $10^{-3}\gamma_0$ are involved in these phenomena \cite{tran2020understanding}), we linearly expand the surface tension of the film according to $\delta$, the thickness of the interface (about $1~n$m), and the global film composition $X=\left(hc+2\Gamma \delta\right)/(h+2\delta)$ around the initial uniform composition $X_0$. Here $\Gamma$ and $c$ are respectively the surface and volume molar fractions of the species with the lower surface tension. We can thus write the surface tension as: $\gamma =\gamma_0\left(1+ \alpha/h-\beta (X-X_0)\right)$. The term $\alpha \gamma_0 /h$ is the Gibbs elastic modulus of the film whereas $\beta$ is a positive dimensionless solutal Marangoni coefficient. If the composition at the interface $\Gamma$ and the one in the bulk $c$ are different, the parabolic flow  $u_{Po}$ contribution to the velocity field will advect differently bulk and interfacial species and the concentration field $X$ will be modified with a flux proportional to $u_{Po}h(\Gamma-c)$. 
In addition, the Marangoni stress controls the velocity gradient at the interface and thus $u_{Po}$ is also found to be proportional to $\gamma_x$. Thus the parabolic velocity field leads to the evolution of $X$ through a term $(\Gamma-c)(h\gamma_x)_x$.
Finally, the relation between the concentration $X$, the height $h$ and the interfacial tension $\gamma$ allows to close the system of equations leading to the evolution equation for $\gamma$. It can be obtained by the time derivative of the expression for interfacial tension in terms of $\delta/h$ and $X-X_0$ written above and using the conservation of species. As detailed in \cite{supplementary}, it is written as:
\begin{equation}\label{eq:3}
  D_t\gamma = \delta\gamma_0\beta(\Gamma_0-c_0)\left(2 \frac{\ubar_x}{h} + \frac{1}{3 \mu} \frac{\left(\gamma_x h\right)_x}{h}\right),
\end{equation}
where $c_0,\Gamma_0$ are the reference bulk and surface concentrations (in molar fractions) of the binary mixture respectively. Note that $\alpha$ is explicitly written as $\alpha=\delta\beta(\Gamma_0-c_0)$.
The first term in the r.h.s. account for the Gibbs' elasticity, while the second term describe the effect of the Marangoni stress.
The equations \eqref{eq:1}, \eqref{eq:2} and \eqref{eq:3} constitute a system of coupled evolution equations necessary to describe the drainage inside a thin film of binary mixture. This system of equations is solved, with appropriate initial and boundary conditions \citep{supplementary}, using a second-order finite difference scheme, with adaptive time-step \citep{duchemin_eggers_2014}. In the limit of vanishing surface tension gradient for pure fluids because in that case $\Gamma_0=c_0$, eq. \eqref{eq:3} becomes redundant and eqs. \eqref{eq:1} and \eqref{eq:2} reduce to a system of equations describing drainage via plug-flow governed only by viscous stretching \citep{breward2002drainage, debregeas1998life}. 
The evolution of the film from an initially prescribed flat interface to a flat profile equilibrated in tension and subsequently to the dimpled shape film thinning is predicted by our reduced-order numerical model and showcased in a supplemental video \citep{supplementary}. 

\begin{figure}
\includegraphics[scale=.29,trim=2mm 4mm 0mm 0mm,clip]{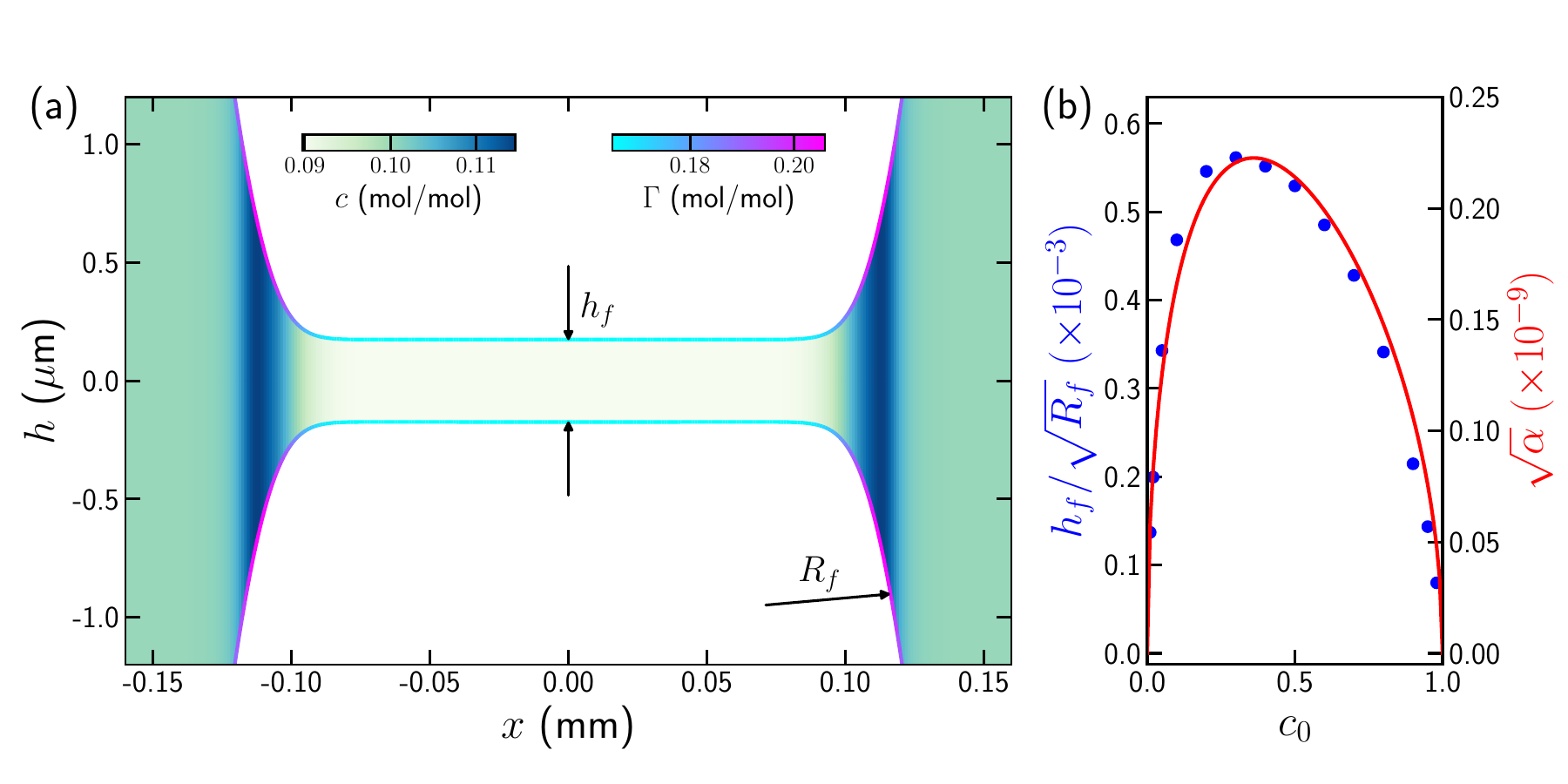}
\caption{\label{fig1} (a) Tension equilibrated film profile and corresponding surface ($\Gamma$) and bulk ($c$) molar fraction distribution for a thin-film of octane-toluene mixture at $t\sim 5\times 10^{-3}$ s. 
(b) Variation of tension equilibrated film thickness, $h_f$ [blue] and the Gibbs parameter, $\alpha=\delta\beta(\Gamma_0-c_0)$ [red] as a function of the reference bulk molar fraction of octane, $c_0=[0.01-0.98]$.} 
\end{figure}

The initial tension relaxation process lasts an inertio-capillary time, $T_c\sim\sqrt{\rho \L^4/\gamma_0\H}$ \cite{lhuissier2012bursting}. This early dynamics ensures mechanical equilibrium in film tension $C$. 
Fig.~\ref{fig1}~(a) shows a typical profile obtained from our model simulation due to an initial stage of tension equilibration. At this moment, the film thickness, $h_f$ and radius at the Plateau border, $R_f$ is plotted in fig.~\ref{fig1}~(b) for the range of mixture concentrations. It satisfactorily matches the analytical scaling: $h_f\sim\sqrt{\alpha R_f}$, previously derived in \cite{tran2022mechanisms}. When a Marangoni stress exists, the drainage time becomes much larger giving a film pinching lifetime, $T_p \gg T_c$ and eq.~\eqref{eq:3} then dominates the film evolution. 

It is important to note that the set of eqs. \eqref{eq:1}~-~\eqref{eq:3} is different from the case of an immobile interface assumption used in Refs. \cite{klaseboer2000film,aradian2001marginal,shah2021influence}.
In contrast to the classical approach of prescribing immobile interfaces and excluding plug-flow \cite{aradian2001marginal}, our model {\amb describes} the leading order plug-flow in the film evolution (eqs. \eqref{eq:1}, \eqref{eq:2}) that equilibrates the film tension. However the pressure is not equilibrated and generates a slow Poiseuille flow. This Poiseuille flow advects differently species at the interfaces and in the bulk, leading to a slow modification of the concentration and thus to the surface tension (eq. \eqref{eq:3}). The surface tension evolution in turn modifies the tension equilibrium. Indeed, the parabolic part contribution to the viscous stress is of $O(h^2)$ compared to the plug flow that relaxes the tension gradient, and is apparently negligible. However, it is {\amb subtly included in the derivation of eq.~\eqref{eq:2} and is the source of non-uniform} advection of species and thus of the Marangoni effect, that leads eventually to pinching.

\begin{figure}
\includegraphics[scale=.43,trim=0mm 73mm 0mm 0mm,clip]{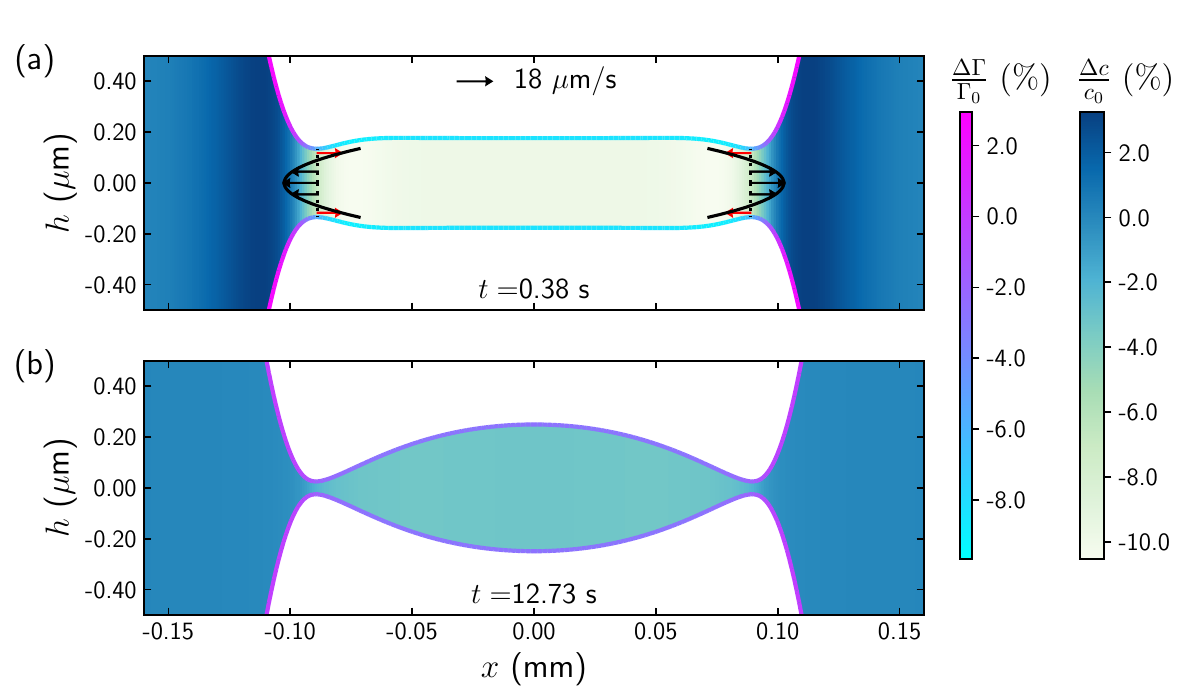}
\caption{\label{fig2} Evolution of marginal pinching in octane-toluene mixture: (a) formation of marginal pinch; (b) thickness of the film reaches $h_c$. The arrows denote the evolved velocity in the film, enveloped by the velocity magnitude at the pinch locations. $\Delta\Gamma=\Gamma-\Gamma_0$ and $\Delta c=c-c_0$, where $\Gamma_0=0.1856$ and $c_0=0.1$ are reference quantities for surface and bulk molar fraction (for octane) respectively.} 
\end{figure}
Fig.~\ref{fig2} shows snapshots for the initial (a) and the final (b) scenarios of the dimple drainage mechanism governed by Poiseuille flow for a typical binary mixture of octane-toluene. Since the surface tension varies inversely with global concentration and with the film thickness, the central part of the film has a higher surface tension. This translates to a Marangoni stress $\sim~\partial_x\gamma$ in the negative direction with respect to capillary drainage. Thus, throughout the film evolution, a Marangoni flux opposes the capillary flow, spontaneously maintaining a parabolic flow profile in the bulk, but with a non zero velocity at the interfaces.

\begin{figure}
\includegraphics[scale=.5,trim=4mm 0mm 0mm 0mm,clip]{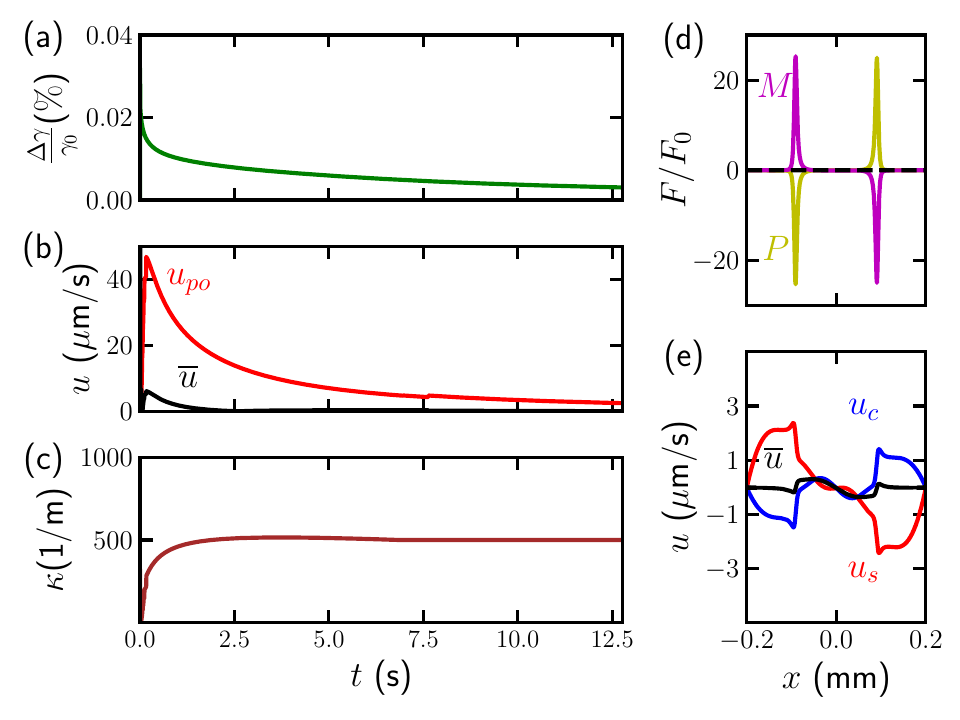}
\caption{\label{fig3} Pinching dynamics of a thin film of Octane-Toluene mixture: evolution of (a) surface tension difference $\Delta\gamma=\gamma_{max}-\gamma_{min}$ across the Plateau border; (b) velocities at the pinch location: mean (plug-flow) velocity $\ubar$ and parabolic (Poiseuille flow) velocity $u_{po}=u^{(2)}h_{min}^2$; (c) maximum curvature, $\kappa$ at the pinch location. Across the film length at $t=12.73$ s: (d) dimensionless force/mass due to Laplace pressure gradient (P) and Marangoni stress (M) where $F_0=1.73\times10^3$ N/Kg. (e) spatial variation of $\ubar$, surface velocity $u_s$, centerline velocity $u_c$.} 
\end{figure}
The driving mechanism for film evolution is the surface tension difference via the second term in the r.h.s of eq.~\eqref{eq:3}. The overall film evolution dynamics thus work towards minimising this difference (plotted as $\Delta\gamma/\gamma_0$ in fig.~\ref{fig3}(a)). Since the pressure gradient is concomitant with the surface tension gradient (through the balance of film tension shown in fig.~\ref{fig3}(d)), it is corollary to say either of the two gradients is minimised during the film evolution. 
Fig.~\ref{fig3}(b) shows the model predictions of the mean (plug-flow) and parabolic (Poiseuille flow) velocities, which asymptotically approach to zero during the remaining process of dimpled drainage. The evolution of curvature at the pinch location, $\kappa=h_{xx}/2$ is shown in Fig.~\ref{fig3}(c). Contrary to the earlier study of \cite{aradian2001marginal}, where they predict a scaling for $\kappa$, we do not observe any scaling for the same, and it essentially remains constant throughout the pinched draining stage. Note that, in the absence of van-der Waals' attractive forces, there is no mechanism for destabilisation and rupture in our model. All the physical quantities like $\Delta\gamma$, $h_{min}$ and $\ubar$ - approach zero asymptotically. Henceforth, to estimate the film lifetime, we choose a cutoff film thickness, $h_c\sim 50~n$m when it is safe to assume that van der Waals force causes spontaneous rupture. A precise discussion of the role of Van der Waals force shows that it marginally modifies the pinching dynamics and that the assumption of rupture occurring when $h=h_c$ is correct \cite{shah2021influence}.
Fig.~\ref{fig3}(e) shows the surface velocity profile across the film length. It is always opposite compared to the capillary drainage velocity ($\ubar$ or $u_{Po}$) during the entire duration of the film evolution.

We now dig deeper into the various mechanisms at play during the film thinning process showcased in fig.~\ref{fig4} and in the supplemental video \cite{supplementary}. At the very shortest timescale, the film tension is equilibrated as explained previously. Since our initial condition is out of this force equilibrium, a quick transient phase can be observed initially. Next, a plug-flow develops causing a \emph{pure stretching} mode which governs the process of equilibration of the film tension, i.e., a surface tension gradient develops which balances the Laplace pressure gradient across the Plateau border. This happens until the inertio-capillary time, $T_c\sim 10^{-2}$ s. Note that this can be simply realised by observing eq.~\eqref{eq:2} where, the source term (in parenthesis) is minimised, damped by the Trouton viscosity. The damped oscillations appearing in fig.~\ref{fig4} are a manifestation of this process. 

\begin{figure}
\includegraphics[scale=.45, trim=0mm 0mm 0mm 12mm, clip]{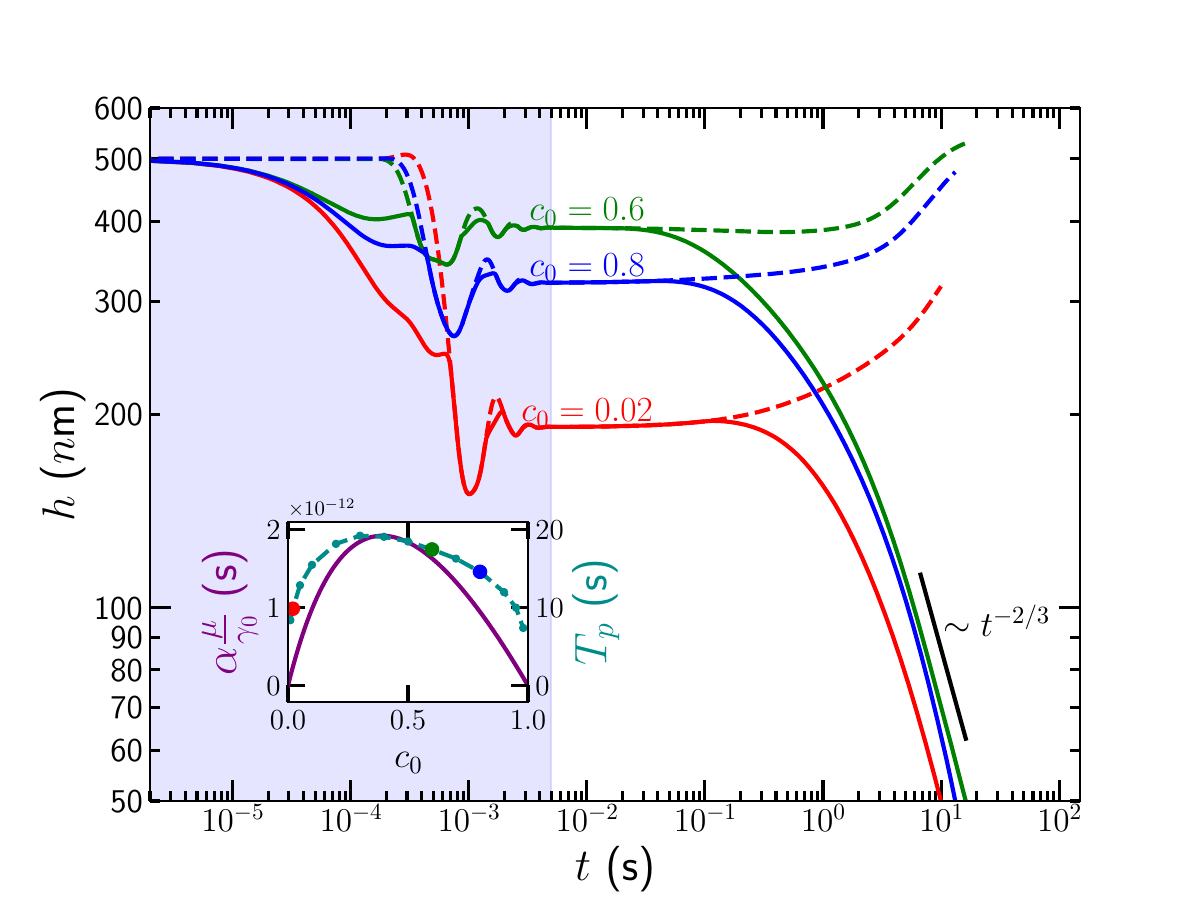}
\caption{\label{fig4} 
Film evolution at the pinch region, $h_{min}$ (solid curves) and at the center, $h_{c}$  (dashed curves) for octane-toluene mixture. The shaded region corresponds to the initial fast process of film tension equilibration. Subsequent relaxation of surface tension leads to the film thinning law: $h_{min}~\sim~t^{-2/3}$. The molar surface, $\sigma=2.5\times 10^5$ m$^2$/mol. [Inset] Model predictions of pinching times of film $T_p$ [teal] compared to $\alpha\mu/\gamma_0$ [purple] for octane-toluene mixture. Color coded circles correspond to $c_0$ values of the main plot.
} 
\end{figure}
From this time onward, the plug-flow is minimised and the surface tension gradient relaxation process comes into play through eq.~\eqref{eq:3}. We thus have a period of transition from a plug-flow to a development of a Poiseuille-flow across the thin film. The duration of this transition phase depends non-monotonically on the mixture concentration. Once the Poiseuille-flow is set up, we have the main stage of film-thinning which follows a power law dependence.
Focusing on the film thinning law, $h_{min}\sim t^{n}$, we observe a value of $n=-0.66\pm 0.06$ throughout the range of mixture concentration, $c_0$. 
In this context, experimental film thinning laws are scarcely reported, but an exponent of $n=-2/3$ has been experimentally found at least in the case of thin-film evolution influenced by sparse surfactants \cite{lhuissier2012bursting}. Our model, being governed by surfactant-like effect predicts close to this scaling, as shown in Fig.~\ref{fig4}. Earlier models \cite{klaseboer2000film,aradian2001marginal,shah2021influence}, for a similar geometry involving Plateau borders, relevant to thin films in foams, predict a scaling exponent of $n=-1/2$ when a Poiseuille flow with immobile interfaces is considered. On the other hand, when a mobile interface is considered with a plug flow in the bulk, the film thins very rapidly with an exponent of $n=-2$ which predicts the dynamics in thin films of pure liquids; as also reported in Ref. \cite{breward2002drainage}. The model presented in this letter does not rely on such assumptions and predicts the film evolution consistent with the experiments of Ref.~\cite{lhuissier2012bursting}. To the best of our knowledge, this is the first time the exponent of $n=-2/3$ for the film thinning law is predicted for pinching of thin-films involving Plateau borders.

Finally, we present our model predictions of film lifetimes in fig.~\ref{fig4} (inset) as a function of mixture concentrations. It has been reported earlier experimentally that the lifetimes of foams and surface bubbles of binary mixture (translated to a length-scale) correlate with the Gibbs modulus parameter, $\alpha$ \citep{tran2022mechanisms}.
Our model predicts a similar correlation and is thus able to account for experimental observations despite the geometrical differences. As a perspective of the current study, we point out that the estimated times are highly sensitive to aspect ratio ($\H/\L$) and length scales ($\L,h_i,R_b$). Further, an estimation of time-scale for the marginal pinching process requires a careful estimation of the dimensions of pinch region, which correlate to the physicochemical parameters of the binary mixture.

To summarize, we develop a model describing dynamics of marginal pinching in surfactant-like systems. The model comprising of coupled evolution equations for film thickness, mean velocity and surface tension describes the interplay between capillary drainage and a concomitant Marangoni flow. The parabolic component of the resulting Poiseuille flow is subtly included in the leading order derivation and additionally appears as a second gradient of surface tension. In the current letter, we obtain variations of pinching times for binary mixtures in a canonical geometry of a Scheludko cell which are in agreement with experiments. Further, the predicted dynamics agree well with qualitative behaviors observed in recent experiments of surface bubbles with sparse-surfactant effects, for example, the \emph{marginal regeneration} reported in contaminated water \citep{lhuissier2012bursting} and the tension equilibration process reported in salt solutions \citep{liu2023nanoscale}.
The model presented here is versatile in nature, providing a recipe to model other foamy systems governed by linear surface tension variation relevant for dilute surfactant-like effects. 
\newpage
\bibliography{Bibliography}

\newpage

\begin{center}
\textbf{\large{\underline {SUPPLEMENTAL INFORMATION}}}
\end{center}

\section{Momentum and mass conservation}\label{sec:model}
For a liquid film with viscosity $\mu$, density $\rho$, and surface tension $\gamma$, the Navier-Stokes equation reads:
\begin{subequations}\label{eq:ns}
\begin{gather}
\dpopa x u + \dpopa z v = 0,\label{eq:nsc}\\
\dpopa t u + u\dpopa x u + v\dpopa z u=-\frac{1}{\rho}\dpopa x P + \frac{\mu}{\rho}\left( \dpopa {xx} u + \dpopa {zz} u \right),\label{eq:nsmx}\\
\dpopa t u + u\dpopa x v + v\dpopa z v=-\frac{1}{\rho}\dpopa z P + \frac{\mu}{\rho}\left( \dpopa {xx} v + \dpopa {zz} v \right),\label{eq:nsmy}
\end{gather}
\end{subequations}
where $P$ is the pressure inside the fluid and $(u,v)$ are velocity components 
along $x$ and $z$ directions respectively.
The above equations hold for $-\frac{h(x,t)}{2}~<~z~<~\frac{h(x,t)}{2}$. At the interfaces ($z=\pm h/2$), the balance of normal and tangential stress is used, the full expression is omitted here for brevity.
Finally, the surface moves with the velocity field at the boundary described by the kinematic equation:
\begin{equation}\label{eq:kin}
    \dpopa t h + u\dpopa x h = \pm 2v.
\end{equation}
Since our geometry involves thin ligament of fluid relative to their elongation, we consider next reducing the order of equations. This approach is similar to \cite{eggers1994drop}, without the gravity and axisymmetric terms.

\subsection{Reduced order equations}
In the lubrication limit, and assuming that the film is flat ($|\partial h/\partial x|\ll 1$), we expand the variables $u,v,P$ in a Taylor series with respect to $z$. Due to symmetry at $z=0$, we take an expansion for $[u,P]$ of the form
\begin{equation}\label{eq:expu}
    [u,P]=[u^{(0)}(x),P^{(0)}(x)] + [u^{(2)}(x),P^{(2)}(x)]\left(z^2-h(x)^2/12\right) + O(z^4).
\end{equation}
This form of expansion gives $u^{(0)}$ as the mean bulk velocity, $\ubar$, at $\mathcal{O}(z^4)$.
Hereafter, we use subscripts to denote derivatives. Using mass conservation \eqref{eq:nsc}, we find~:
\begin{equation}\label{eq:expv}
        v=-\ubar_x z - u_x^{(2)}\left(z^3/3-h^2z/12\right) + u^{(2)}(hh_xz/6) + O(z^5).
\end{equation}

Inserting \eqref{eq:expu}-\eqref{eq:expv} into \eqref{eq:nsmx}-\eqref{eq:kin}, we reduce the equations to the leading order in $z$. The $x$-momentum equation \eqref{eq:nsmx} becomes
\begin{equation}\label{eq:nsmx0}
    \ubar_t + \ubar~\ubar_x = - \frac{P_x^{(0)}}{\rho} + \frac{\mu}{\rho} (\ubar_{x
x} + 2 u ^{(2)}),
\end{equation}
and $z$-momentum equation \eqref{eq:nsmy} is identically satisfied. 
Considering only the upper interface ($z=h/2$), the normal and tangential interfacial stress conditions and the kinematic boundary condition \eqref{eq:kin} reduce respectively to~:
\begin{subequations}\label{eq:jumpbc0}
\begin{align}
    \frac{P ^{(0)}}{\rho} = & - \frac{\gamma h_{xx}}{2\rho} - 2 \frac{\mu}{\rho} \ubar_x,\label{eq:nsbc0}\\
    u ^{(2)} h = &\frac{\gamma_x}{\mu} + \frac{\ubar_{x x} h}{2} + 2 h_x \ubar_x,\label{eq:tsbc0}\\
    h_t = &-(\ubar h)_x.\label{eq:hkin}
\end{align}
\end{subequations}
Note that the leading order tangential stress condition \eqref{eq:tsbc0} is at $O(z)$. Next, we eliminate $P^{(0)}$ and $u^{(2)}$ in \eqref{eq:nsmx0} using \eqref{eq:nsbc0} and \eqref{eq:tsbc0} respectively, finally obtaining
\begin{equation}\label{eq:uevol0}
\ubar_t + \ubar~\ubar_x = \left( \frac{(h_{x x} \gamma)_x}{2 \rho} + 2 \frac{\gamma_x}{\rho h} \right) + \frac{\mu}{\rho}\frac{(4h \ubar_x)_x}{h}.
\end{equation}
The first term in the r.h.s. (in parenthesis) which comprises the Laplace pressure and Marangoni stress is simplified further based on intuition. First, $h h_{xx}\gamma_x/2$ can be neglected in comparison to $2\gamma_x$ as $h\ll x$. Second, denoting a reference surface tension of the mixture at a given mixture concentration and infinite film thickness as $\gamma_0$, we can assume $\gamma h_{xxx}\simeq \gamma_0 h_{xxx}$ as the variations in $\gamma$ are considered to be small ($\sim O(h)$). 

\section{Species transport}
We derive an evolution equation for a component of binary mixture based on a volume conservation law. Let $c_1$ be the volume fraction of species (with the lower surface tension) in the bulk and $\Gamma_1$ be the corresponding volume fraction in the surface layer. Considering a finite thickness of the interface, $\delta$, the total molar fraction of species 1 in  a thin differential element in $x$, can be written as $w_1=hc_1+2\delta\Gamma_1$. The flux is expressed as $j_1 = c_1 h \ubar + \Gamma_1 (2\delta)u_s$, where $\ubar$ and $u_s$ are mean bulk and interfacial velocities respectively. Omitting the index 1, and using subscripts to denote derivatives hereafter, the conservation of molar fraction os species 1 is written as $w_t = -j_x = -(c h \overline{u} + \Gamma (2\delta)u_s)_x$. The interfacial velocity can be evaluated using \eqref{eq:expu} giving $u_s=\ubar+u^{(2)}h^2/6$. We thus obtain: 
\begin{equation}
    w_t = -\left[h c \ubar + 2 \delta \Gamma \left( \ubar + \frac{u^{(2)} h^2}{6} \right)\right]_x.
\end{equation}
Substituting \eqref{eq:tsbc0} for $u^{(2)}$ in the equation above and assuming the surface tension gradient term, $\gamma_x/\mu h$ to be the dominant term in the expression as explained below \eqref{eq:uevol0}, we obtain:  
\begin{equation}
    w_t = - (\ubar w)_x - \frac{\delta}{3 \mu} [\gamma_x h \Gamma]_x .
\label{eq:w}
\end{equation}
Note that although the Poiseuille flow (second term in the r.h.s) is apparently negligible, we retain it in the leading order equation to include its effect on the species advection, as discussed in the letter.
We thus have three evolution equations for $h$ \eqref{eq:hkin}, $\ubar$ \eqref{eq:uevol0} and $w$ \eqref{eq:w} but two additional unknowns, $\gamma,\Gamma$. However these two variables are intrinsically connected to $w$ following thermodynamics of ideal solutions. In the next section, we explore the relations among these variables 
to derive an evolution equation for the surface tension $\gamma$.  
\section{Surface tension of binary mixtures}\label{sec:st}
Thermodynamics of ideal mixtures leads to a relation between the surface molar fraction, $\Gamma_i$, and the bulk molar fraction $c_i$ for an infinite reservoir known as Butler's equation \citep{butler_thermodynamics_1932}. Considering the same molar surface for the two species ($\sigma_i=S.RT$), where $RT$ is the molar thermal energy, we have: 
\begin{equation}
  \Gamma_i=c_i e^{(S (\gamma -\gamma_i))}, \label{eq:Gamma}
\end{equation}
where $\gamma$ is the effective surface tension of the mixture, $\gamma_i$ the surface tension of pure liquid species $i$. For a two species mixture, we have $\Gamma_1+\Gamma_2=1$ and $c_1+c_2=1$. This gives the interfacial tension for an infinite reservoir,
\begin{equation}
\gamma(c_1)= \gamma_2-  \frac{\log \left[ 1+c_1(e^{S(\gamma_2-\gamma_1)}-1)\right]}{S}.
 \label{eq:gamma_inf}
\end{equation}
Note that $\gamma=\gamma_2$ for $c_1=0$ (pure species 2) and $\gamma=\gamma_1$ for $c_1=1$ (pure species 1). Hereafter, we omit the index $1$ for $c$ and $\Gamma$ for brevity and use the index $0$ for reference values at infinite film thickness. Next, we define a new variable, the global molar fraction, 
\begin{equation}\label{eq:Xdef}
 X =
\frac{w}{h+2\delta} = \frac{c h + 2 \Gamma \delta}{h+2\delta}.   
\end{equation}
Subsequently, an explicit relation between $X$ and $\gamma$ can be deduced, giving the expression: 
\begin{equation}
   \gamma= \gamma_2 -\frac{1}{S} \log \left(\frac{X (2 \delta +h) e^{S(\gamma_2 -\gamma_1)}}{h+2 \delta  e^{S (\gamma -\gamma_1)}}-\frac{X (2 \delta +h)}{h+2 \delta  e^{S (\gamma -\gamma_1)}}+1\right).
\end{equation}
Next, substituting $\gamma_0=\gamma(c_0)$ for $\gamma$ in the RHS using \eqref{eq:gamma_inf}, and expanding it in Taylor series both in $\delta$ and $(X - X_0)$ - $X_0$ being the initial uniform molar fraction - gives the following - exact - result at the first order in $\delta$ and $X-X_0$:
\begin{equation}
  \gamma  =  \gamma_0+\frac{\delta}{h}\frac{2 (1-X_0) X_0 \left(e^{S\Delta\gamma}-1\right)^2}{ S \left(1+X_0(e^{S\Delta\gamma}-1)\right)^2} -(X-X_0)\frac{e^{S\Delta\gamma}-1}{S(1+ X_0 \left(e^{S\Delta\gamma}-1\right))},
\end{equation}
where $\Delta\gamma=\gamma_2-\gamma_1$. We have thus the linearised version of $\gamma$:
\begin{equation}  \label{eq:gamma}
  \gamma = \gamma_0\left(1 + \frac{\alpha}{h} - \beta \left(X-X_0 \right)\right).
\end{equation}
Combining \eqref{eq:Gamma} and \eqref{eq:gamma_inf}, and using \eqref{eq:Xdef} we also have:
\begin{equation}
\Gamma_0-c_0=\frac{(1-X_0) X_0 \left(e^{S\Delta \gamma}-1\right)}{1 + X_0 \left(e^{S\Delta \gamma}-1\right)},
  \label{Gamma-c}
\end{equation}
which leads to,
\begin{equation} 
    \beta=\frac{1}{\gamma_0 S} \frac{\Gamma_0-c_0}{X_0(1-X_0)}=-\frac{1}{\gamma_0}\frac{\partial \gamma}{\partial c_0},
    \label{beta}
\end{equation}
and to,
\begin{equation}
\alpha=2\delta (\Gamma_0-c_0)\beta.
\label{alpha}    
\end{equation}

\begin{figure}
\includegraphics[scale=.45,trim=0mm 0mm 0mm 0mm,clip]{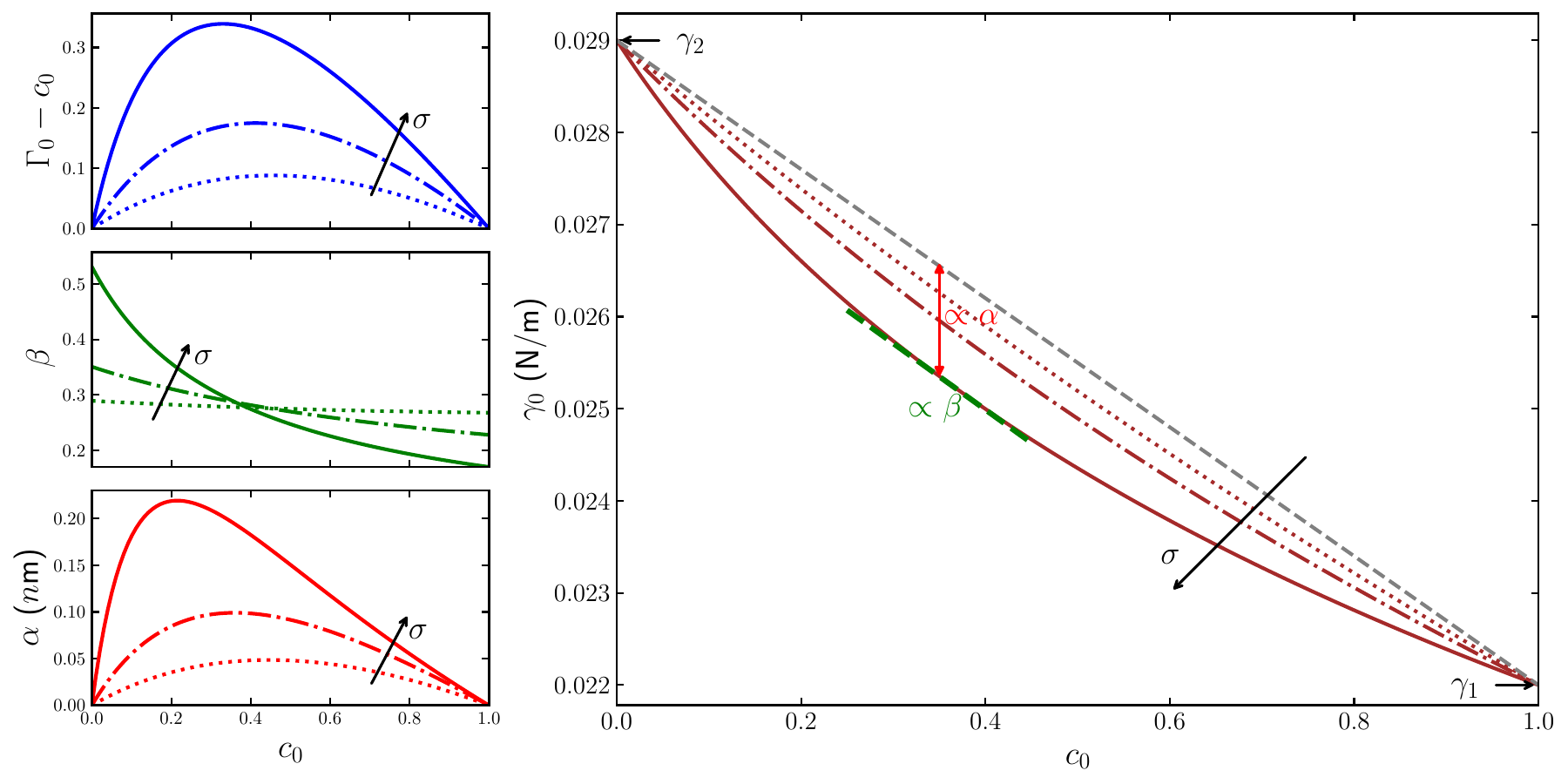}
\caption{\label{phychem} Reference parameters of an octane-toluene mixture. Variation of $(\Gamma_0-c_0)$, $\beta$, $\alpha$ and $\gamma_0$ with reference concentration, $c_0$ for three values of molar surface $\sigma=$ $1.25\times10^5$ (dotted), $2.5\times10^5$ (dot-dashed) and $5\times10^5$ (solid) m$^2/$mol.
}
\end{figure}
The expression in parenthesis of eq.~\eqref{eq:gamma} is a dimensionless quantity which is in practice $O(1)$ and is always positive, with the choice that species $1$ has the lowest interfacial tension. The second and third terms are the thickness-dependent and concentration-dependent corrections respectively which are generally at most $O(\delta/h)$. Note that $\alpha$ vanishes both for $X_0=0$ and $X_0=1$ that corresponds to the case of pure liquids, where there is no Marangoni effect. Fig.~\ref{phychem} shows the variation of these parameters as a function of bulk concentration, $c_0$ and molar surface, $\sigma$. In a binary mixture, the surface tension varies sub-linearly with bulk concentration \citep{tran2020understanding}. As can be observed in the right panel, $\alpha$ and $\beta$ characterises this sub-linearity in terms of extent and slope respectively.

We now evaluate the derivative of \eqref{eq:gamma},
obtaining, 
\begin{equation}
    \gamma_t=-\alpha\gamma_0 \frac{h_t}{h^2}-\beta\gamma_0 X_t.
    \label{eq:gamma_t}
\end{equation}
To evaluate $X_t$, we use \eqref{eq:Xdef} and easily obtain,
\begin{equation}
    X_t=\frac{w_t}{h+2 \delta}-\frac{w h_t}{(h+2 \delta)^2}.
\end{equation}
Conservation of species gives $w_t=-\left(\Bar{u}hc+u_s2\delta \Gamma\right)_x$
and volume conservation gives 
$h_t=-\left(\Bar{u}h+2u_s\delta\right)_x$.
Using these two laws along with the definition, $w=hc+2\delta\Gamma$ and the derivative $c_x$ evaluated from \eqref{eq:Xdef}, we get an evolution equation for global molar fraction (in the limit of $\delta\ll h$),
\begin{equation}
     X_t=-\Bar{u}X_x  + 
    (\Gamma-c) \frac{2\delta}{h}\left[\Bar{u}-u_s\right]_x .
\end{equation}
Injecting the above expression in \eqref{eq:gamma_t}, back-substituting for $X_x,h_t,u_s$ and using \eqref{beta} and \eqref{alpha}; we subsequently obtain:  
\begin{equation}\label{eq:gevol}
  D_t\gamma = \delta\gamma_0\beta(\Gamma_0-c_0)\left(2 \frac{\ubar_x}{h} + \frac{1}{3 \mu} \frac{\left(\gamma_x h\right)_x}{h}\right).
\end{equation}

The equations \eqref{eq:gevol}, \eqref{eq:uevol0} and \eqref{eq:hkin} constitute a closed-form coupled evolution equations necessary to describe the drainage inside a thin film of binary mixture with mobile interfaces. For pure fluids, we have $\Gamma_0=0=c_0$,and \eqref{eq:gevol} becomes redundant and \eqref{eq:uevol0} and \eqref{eq:hkin} reduces to the system of equations described in \cite{breward2002drainage} for pure fluids governed only by viscous stretching. Additionally imposing immobile interface translates to using a different expansion for the parabolic velocity, $u=u^{(0)}\left(z^2 - h^{2}/4\right)+O(z^4)$. This gives a single evolution equation for film height: $h_t=-\gamma_0(h_{xxx}h^3)_x/24\mu$, the same form as discussed in \cite{aradian2001marginal}, relevant in the case of concentrated surfactant condition.

\section{Direct numerical simulation}\label{sec:dns}
For the scope of this study, we consider a ligament of fluid pinned between two walls at a distance $\mathcal{L}$ and having a fixed half thickness $\E\L$ at the wall, where $\E$ is an aspect ratio. This is close to Scheludko cell experiments but with an added constraint of confined drainage. However, the pinching dynamics is a localised phenomenon and is expected not to be significantly affected by the far-field conditions. We choose an arbitrary time-scale based on capillary emptying defined as $\mathcal{T}=\sqrt{\rho L^3/\gamma_0}$. The evolution equations \eqref{eq:hkin}, \eqref{eq:uevol0} and \eqref{eq:gevol} can be transformed into the dimensionless form:
\begin{subequations}\label{eq:gov}
\begin{align}
    H_T= &-(\Ubar H)_{\chi},\label{eq:gov1}\\
    \Ubar_T= &\left(\frac{H_{\chi\chi\chi}}{2}+2\frac{G_{\chi}}{H}\right) + 4 \Oh \left(\Ubar_{\chi\chi}+\frac{H_{\chi} \Ubar_{\chi}}{H}\right),\label{eq:gov2}\\
    G_T= & \Ma\frac{\Ubar_{\chi}}{H} - \Ubar G_{\chi} + \frac{\Mb(\Gamma_0-c_0)}{3\Oh}\left(G_{\chi\chi}+\frac{H_{\chi} G_{\chi}}{H}\right).\label{eq:gov3}
\end{align}
\end{subequations}
Here, all lengths and time are scaled by $\mathcal{L}$ and $\mathcal{T}$ respectively, using
$\chi=x/\mathcal{L}$, $H=h/\mathcal{L}$, $T=t/\mathcal{T}$. Further, the velocity is scaled as $\ubar=\Ubar\mathcal{L}/\mathcal{T}$ and surface tension as $\gamma=G\gamma_0$. We use the dimensionless numbers: Ohnesorge number, $\Oh=\sqrt{\mu^2/\rho\gamma_0\mathcal{L}}$; Gibbs number $\Ma=\alpha/\mathcal{L}$ and Marangoni parameter $\Mb=\beta D$ where $D=\delta/\mathcal{L}$. 

The initial conditions for the variables are taken as $G=1$ and $\Ubar=0$. For $H$, we prescribe a film profile described by a Plateau border of constant radius $R_b$ connected to a flat film at the center \citep{shah2021influence}:
\begin{align}
    H = & \frac{h_i/2}{\L} & \text{at}~\left(-\L/4 < x < \L/4\right),\\
    H = & \frac{h_i/2}{\L} + \left\{\frac{x^2 - (\L/4)^2}{R_b} + \frac{2\left(\L/4\right)^2}{R_b}\ln{\left(\frac{\L/4}{x}\right)}\right\}/\L & \text{at}~\left(\L/4 < x < \L, -\L < x < -\L/4 \right).
\end{align}

\begin{figure}
\includegraphics[scale=.55,trim=0mm 5mm 0mm 0mm,clip]{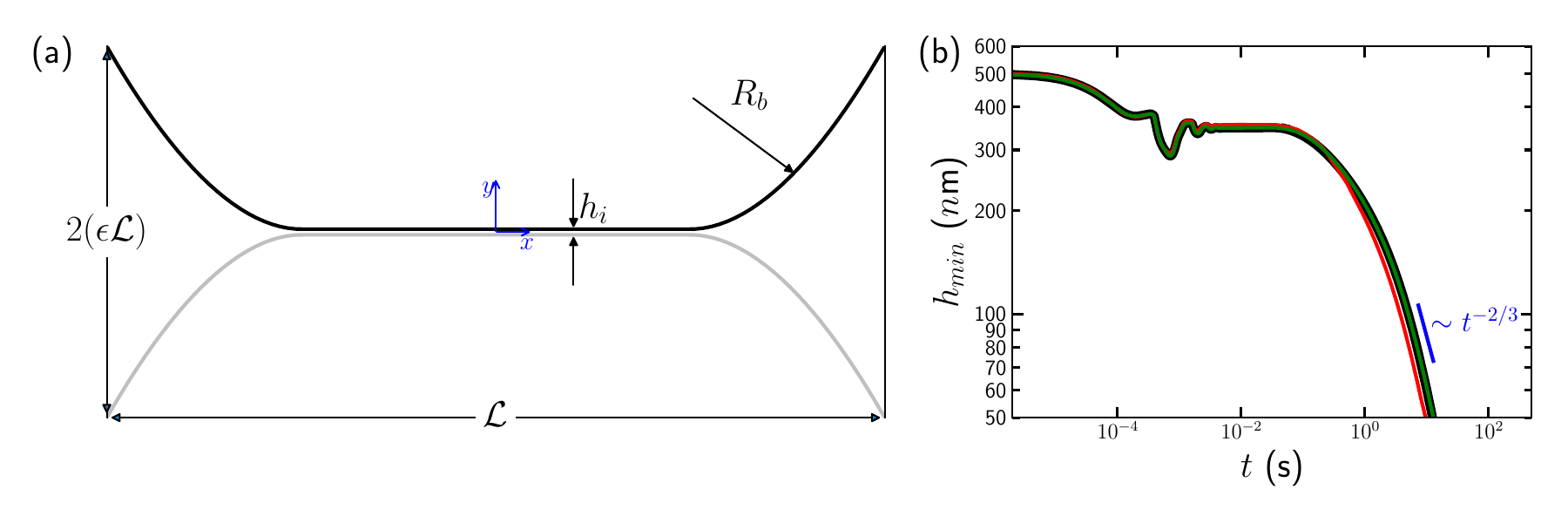}
\caption{\label{schem}. (a) Schematic of the initial film profile and computational domain. The numerical code solves for the black solid line. The grey line is a mirror image across $y=0$. (b) Evolution of film thickness at the pinching point $h_{min}$. Red: $O(\Delta T)$ accuracy, $\Delta T < 5\times10^{-3}, \Delta \chi = 1/100, t_{s}=55$ min. Black: $O(\Delta T^2)$ accuracy, $\Delta T = 1\times10^{-3}, \Delta \chi = 1/400, t_{s}=520$ min. Green: $O(\Delta T^2)$ accuracy, $\Delta T < 10^{-2}, \Delta \chi = 1/200, t_{s}=110$ min. Here, $t_s$ is the simulation time done on an Apple M1 (2020) processor.
}
\end{figure}

The parameters are described in fig.~\ref{schem}. Using experimental estimations reported in \cite{tran2022mechanisms}, we prescribe $h_i=500~n$m, $\L=0.4$ mm, $R_b=1$ mm and $\delta=1~n$m. With these constraints, we obtain an aspect ratio, $\E=4.1\times 10^{-2}$ $D=2.5\times 10^{-6}$. $M_{\alpha}$ and $M_{\beta}$ are fixed by $\alpha,\beta$ which are only a function of $c_0$ and $\sigma$. In all the simulation data reported, $\sigma=2.5\times10^5$ m$^2/$mol.

%

For boundary conditions, we have Dirichlet conditions $H=H_0$ and $\Ubar=0$ at the two boundaries, whereas a homogeneous Neumann condition is used for $G$. Note that $G_{\chi}=0$ also ensures the additional boundary condition, $H_{\chi\chi\chi}=0$ required for $H$ through the tension equilibration condition, $HH_{\chi\chi\chi}=4G_{\chi}$ (see the letter for more information about tension equilibration).

The set of evolution equations \eqref{eq:gov1}-\eqref{eq:gov3} is solved by using a semi-implicit finite difference scheme on a uniform grid. All the higher derivatives of $H,\Ubar$ and $G$ are discretised implicitly using central finite difference with second order accuracy with a prescribed resolution of $N=400$. The time-integration is done by a first order accurate Euler forward difference, extrapolated to second order accuracy using the Richardson extrapolation scheme \citep{duchemin_eggers_2014}. Further, an adaptive time-stepping scheme is used based on an error criterion of $E=$ max$(|H_2 - H_1|)/\E < 10^{-6}$. Here, $H_1=H(\Delta T), H_2=H(2\times\Delta T/2)$. The code is written in \texttt{python} using the \emph{numpy} library and the discretised semi-implicit linear equations are solved using \emph {linalg.solve} function at a given time-step. Fig.~\ref{schem}(b) shows the accuracy of the numerical scheme with some relevant data mentioned in the caption. For all the simulations reported in the letter, we choose the scheme corresponding to the green curve for an optimal accuracy and computational time.

%


\end{document}